\documentclass[aps, prd, amsmath, floats, floatfix, twocolumn, nofootinbib, superscriptaddress]{revtex4-1}
\usepackage{graphicx}
\usepackage{color}
\usepackage{latexsym}

\newcommand{\be}{\begin{eqnarray}}
\newcommand{\ee}{\end{eqnarray}}

\begin{document}

\title{Note on a new parametrization for testing the Kerr metric}

\author{M. Ghasemi-Nodehi}
\affiliation{Center for Field Theory and Particle Physics and Department of Physics, Fudan University, 200433 Shanghai, China}

\author{Cosimo Bambi}
\email[Corresponding author: ]{bambi@fudan.edu.cn}
\affiliation{Center for Field Theory and Particle Physics and Department of Physics, Fudan University, 200433 Shanghai, China}
\affiliation{Theoretical Astrophysics, Eberhard-Karls Universit\"at T\"ubingen, 72076 T\"ubingen, Germany}

\date{\today}

\begin{abstract}
We propose a new parametrization for testing the Kerr nature of astrophysical black hole candidates. The common approaches focus on the attempt to constrain possible deviations from the Kerr solution described by new terms in the metric. Here we adopt a different perspective. The mass and the spin of a black hole make the spacetime curved and we want to check whether they do it with the strength predicted by general relativity. As an example, we apply our parametrization to the black hole shadow, an observation that may be possible in a not too distant future.
\end{abstract}

\maketitle


\section{Introduction}

General relativity has been tested in weak gravitational fields by experiments in the Solar System and observations of binary pulsars~\cite{will}. The interest has recently shifted to testing the theory in more extreme conditions. In this context, an important line of research is devoted to verify the Kerr nature of astrophysical black hole candidates~\cite{r1,r2,r3,r4,cc1,cc1.5,cc2,cc3,cc4}. In the framework of general relativity, the spacetime geometry around astrophysical black holes should be well described by the Kerr solution. Any observation of deviation from the Kerr metric should thus be interpreted as evidence of new physics.

If we want to test the Schwarzschild metric in the weak field regime, we can write the most general static, spherically symmetric, and asymptotically flat metric with an expansion in $M/r$, where $M$ is the mass of the central object and $r$ is some radial coordinate. The approach is traditionally formulated in isotropic coordinates and the line element reads
\be\label{eq-ppn0}
ds^2 &=& - \left( 1 - \frac{2 M}{r} + \beta \frac{2 M^2}{r^2} + ... \right) dt^2 
\nonumber\\ &&
+ \left( 1 + \gamma \frac{2 M}{r} + ... \right) \left(dx^2 + dy^2 + dz^2\right) \, .
\ee
For $M=0$, we have a flat spacetime. The term $-2M/r$ in $g_{tt}$ is to recover the correct Newtonian limit. Since there are not other natural constraints, the coefficients in front of higher orders terms are {\it a priori} unknown and parameterized by $\beta$ and $\gamma$, which must be measured by observations. Current observations require~\cite{ppn1,ppn2}
\be\label{eq-ppn}
|\beta - 1| < 2.3 \cdot 10^{-4} \, , \quad
|\gamma - 1| < 2.3 \cdot 10^{-5} \, .
\ee 
When we write the Schwarzschild solution in isotropic coordinates, we find that $\beta = \gamma = 1$. Current observations thus confirm the Schwarzschild metric in the weak field regime at the level of precision accessible with the available facilities.

In the case of tests in the strong field regime, there are some complications. In particular, it is not possible to perform an expansion in $M/r$, because this is not a small parameter any more. This fact leads to have an arbitrary number of possible deviations from the Kerr metric and it is impossible to perform a completely model-independent analysis as in the case of weak field tests in the Solar System. The common approach employed in the past few years is to adopt as metric an ansatz with a number of ``deformation parameters''. This metric reduces to the Kerr solution if all the deformation parameters vanish. Deviations from the Kerr geometry appear in the case of non-vanishing deformation parameters. Like the $\beta$ and $\gamma$ parameters in Eq.~(\ref{eq-ppn0}), the deformation parameters are supposed to be constants to be measured by observations. The latter can confirm the Kerr nature of black hole candidates by constraining the values of the deformation parameters. There are several proposals in the literature, each of these with its advantages and disadvantages~\cite{t1,t2,t3,t4,t5,t6}.

In this paper, we explore a slightly different approach. In the metrics proposed in~\cite{t1,t2,t3,t4,t5,t6}, the Kerr solution is recovered when all the deformation parameters vanish. The deformation parameters introduce thus some additional terms, which should describe the spacetime geometry around black holes in alternative theories of gravity. The spirit of the parametrization proposed here is to check how the mass and the spin angular momentum make the spacetime curved. In general relativity, we exactly know how the mass and the spin angular momentum of a black hole deform the geometry of the spacetime. Geodesic motion is completely determined by the Kerr metric. Observations can thus confirm the predictions of general relativity if the mass and the spin terms are exactly where they are and with the correct coefficient as they appear in the Kerr metric. We introduce 11 ``Kerr parameters'', which are all equal 1 in the Kerr metric. If these parameters are larger or smaller than 1, the mass and the spin deform the spacetime geometry more or less than what it is predicted by general relativity. We would like to stress that, strictly speaking, our proposal is simply a different parametrization to test the Kerr metric. However, the spirit is different because we try to verify the Kerr solution rather than the possible presence of new terms.

In order to illustrate advantages and disadvantages of our proposal, we apply our parametrization to the black hole shadow, a kind of observation that may be possible in a not too distant future. Now we can understand which parts of the Kerr metric can be tested by observations and which parts cannot be verified because do not leave any observational signature.

In what follows, we will employ natural units in which $G_{\rm N} = c = 1$ and adopt a metric with signature $(-+++)$.

\section{Parameterization of the Kerr metric}

In Boyer-Lindquist coordinates, the line element of the Kerr metric reads
\begin{widetext}
\be\label{eq-k}
ds^2 &=& - \left( 1 - \frac{2 M r}{r^2 + a^2 \cos^2\theta} \right) dt^2 
- \frac{4 M a r \sin^2\theta}{r^2 + a^2 \cos^2\theta} dt d\phi 
+ \frac{r^2 + a^2 \cos^2\theta}{r^2 - 2 M r + a^2} dr^2 
\nonumber\\ &&
+ \left( r^2 + a^2 \cos^2\theta \right) d\theta^2 
+ \left( r^2 + a^2 + \frac{2 M a^2 r 
\sin^2\theta}{r^2 + a^2 \cos^2\theta} \right) \sin^2\theta d\phi^2 \, ,
\ee
\end{widetext}
where $M$ is the black hole mass and $a$ is the black hole spin parameter. $|a| \le M$ is the condition for the existence of the event horizon. For $|a| > M$, there is no horizon and the Kerr metric describes the spacetime with a naked singularity.

For $M = a = 0$, Eq.~(\ref{eq-k}) reduces to the line element of the flat spacetime in spherical coordinates. Since we want to test how each $M$ and $a$ term appearing in Eq.~(\ref{eq-k}) deform the spacetime geometry, we rewrite Eq.~(\ref{eq-k}) by introducing 11~parameters $b_i$ as follows
\begin{widetext}
\be\label{eq-proposal}
ds^2 &=& - \left( 1 - \frac{2 b_1 M r}{r^2 + b_2 a^2 \cos^2\theta} \right) dt^2 
- \frac{4 b_3 M a r \sin^2\theta}{r^2 + b_4 a^2 \cos^2\theta} dt d\phi 
+ \frac{r^2 + b_5 a^2 \cos^2\theta}{r^2 - 2 b_6 M r + b_7 a^2} dr^2 
\nonumber\\ &&
+ \left( r^2 + b_8 a^2 \cos^2\theta \right) d\theta^2 
+ \left( r^2 + b_9 a^2 + \frac{2 b_{10} M a^2 r 
\sin^2\theta}{r^2 + b_{11} a^2 \cos^2\theta} \right) \sin^2\theta d\phi^2 \, .
\ee
\end{widetext}
In the case of the Kerr metric, $b_i = 1$ for all $i$. If one of these coefficients were larger (smaller) than 1, the interpretation would be that the associated mass or spin term distorts the spacetime geometry more (less) than what predicted by general relativity.

Spacetime curvature is associated to the Riemann tensor. If the Riemann tensor vanishes in a coordinate system, it vanishes in any coordinate system and the spacetime is flat. The metric is instead a quantity strongly related to the coordinate system, which is arbitrary. The expressions of curvature invariants (e.g. the Kretschmann scalar $\mathcal{K} = R_{\mu\nu\rho\sigma} R^{\mu\nu\rho\sigma}$) of the metric in~(\ref{eq-proposal}) are extremely long and do not provide any particular insight, so they are not reported here. In order to motivate our parametrization to test whether the mass and the spin make the spacetime curved as predicted by general relativity, we can simply note that, if we indicate with $\mathcal{K}^{\rm Kerr}$ the Kretschmann scalar of the Kerr metric, for our metric with $\{ b_i \}$ we have
\be\label{eq-vbv}
\mathcal{K} = \mathcal{K}^{\rm Kerr} + 
\sum_{i=1}^{11} \mathcal{K}^{(1)}_i X_i \left( b_i - 1 \right) + ... \, ,
\ee
where $X_i = M$, $a$, $Ma$, $a^2$ or $Ma^2$ depending on $i$. Eq.~(\ref{eq-vbv}) is the expansion of the Kretschmann scalar around $b_i = 1$. If some $b_i \neq 1$, the mass and/or the spin deform the spacetime in a different way.

It is worth noting that $b_1$, $b_3$, and $b_6$ are already constrained by observations. In the case of $b_1$ and $b_6$, this can be easily seen by rewriting the line element in Eq.~(\ref{eq-ppn0}) in Schwarzschild-like coordinates
\be
ds^2 &=& - \left[ 1 - \frac{2 M}{r} + \left( \beta - \gamma \right) \frac{2 M^2}{r^2} + ... \right] dt^2 
\nonumber\\ &&
+ \left( 1 + \gamma \frac{2 M}{r} + ... \right) dr^2 
\nonumber\\ &&
+ r^2 d\theta^2 
+ r^2 \sin^2\theta d\phi^2 \, .
\ee
$b_1 = 1$ in order to recover the Newtonian limit. $b_6 \approx 1$ with a precision of $10^{-5}$ to satisfy the Solar System constraints in Eq.~(\ref{eq-ppn}). The Lense-Thirring effect has been confirmed with a precision of 10\% from the study of Earth-orbiting satellites~\cite{lte1,lte2}. This implies $b_3 \approx 1$ at the level of 10\%. The other 8 parameters ($b_2$, $b_4$, $b_5$, $b_7$, $b_8$, $b_9$, $b_{10}$, $b_{11}$) are currently not constrained by observations.

\section{Black hole shadow}

The ``shadow'' of a black hole is a dark region over a brighter background appearing in the direct image of an accreting black hole. If the accretion flow is an optically thin emitting medium surrounding the compact object, the boundary of the shadow corresponds to the photon capture sphere as seen by a distant observer~\cite{falcke}. In the case of a geometrically thin and optically thick accretion disk, the boundary of the shadow corresponds to the apparent image of the inner edge of the disk~\cite{fukue,cfm}.

In the past years, there have been a significant work to calculate the shadows of black holes in general relativity and in alternative theories of gravity~\cite{sh0,sh1,sh2,sh3,sh4,sh5,sh6,sh7,sh8,sh11,sh12}. Most studies have focused on the calculation of the apparent shape of the photon capture sphere, namely the case of black holes surrounded by optically thin emitting medium. The interest on this topic is in part motivated by the possibility of detecting in the future the shadow of SgrA$^*$, the supermassive black hole candidate at the center of the Galaxy, with sub-mm very long baseline interferometry (VLBI) facilities~\cite{doe}. At sub-mm wavelength, the medium around SgrA$^*$ should become optically thin, the interstellar scattering should be significantly reduced, and interferometric techniques at sub-mm wavelength should be able to reach a resolution comparable to the gravitational radius of this object.

As suggested by its name, the photon capture sphere is a 2-surface in the 3-space defining the boundary between the photon that, arriving from a certain point at infinity, are captured by the black hole or scattered back to infinity. If a photon enters the photon capture sphere, eventually it is lost because swallowed by the black hole. If a photon does not enter the photon capture sphere, eventually it comes back to infinity.

We calculate the boundary of the photon capture sphere as seen by a distant observer as we have done in Ref.~\cite{noi}. We consider the image plane of the distant observer with Cartesian coordinates $(X,Y)$. We fire photons from a certain grid of the image plane. The photon initial conditions are completely determined by the position of the photon in the grid, because the photon 3-momentum must be perpendicular to the image plane. If the photon has Cartesian coordinates $(X_0,Y_0)$, the initial conditions with respect to the Boyer-Lindquist coordinates are~\cite{cfm}
\be\label{eq-1}
t_0 &=& 0 \, , \nonumber\\
r_0 &=& \sqrt{X^2_0 + Y_0^2 + D^2} \, , \nonumber\\
\theta_0 &=& \arccos \frac{Y_0 \sin i + D \cos i}{\sqrt{X_0^2 + Y_0^2 + D^2}} \, , \nonumber\\
\phi_0 &=& \arctan \frac{X_0}{D \sin i - Y_0 \cos i} \, .
\ee
and the 4-momentum of the photon is
\be\label{eq-2}
k^r_0 &=& - \frac{D}{\sqrt{X_0^2 + Y_0^2 + D^2}} |\bf{k}_0| \, , \nonumber\\
k^\theta_0 &=& \frac{\cos i - D \frac{Y_0 \sin i + D 
\cos i}{X_0^2 + Y_0^2 + D^2}}{\sqrt{X_0^2 + (D \sin i - Y_0 \cos i)^2}} |\bf{k}_0| \, , \nonumber\\
k^\phi_0 &=& \frac{X_0 \sin i}{X_0^2 + (D \sin i - Y_0 \cos i)^2} |\bf{k}_0| \, , \nonumber\\
k^t_0 &=& \sqrt{\left(k^r_0\right)^2 + r^2_0  \left(k^\theta_0\right)^2
+ r_0^2 \sin^2\theta_0  (k^\phi_0)^2} \, , 
\ee
where $D$ is the radial coordinate of the distant observer and $i$ is the angle between the black hole spin and the line of sight of the distant observer. In our calculations, $D = 10^6$~$M$, which is far enough to assume that the background geometry is flat. $k^t_0$ is thus obtained from the condition $g_{\mu\nu}k^\mu k^\nu = 0$ with the metric tensor of a flat spacetime.

The boundary of the apparent photon capture sphere is the closed curve in the image plane of the distant observer separating the photons that are capture by the black hole from those that approach the black hole but are eventually scattered to infinity. In the case of an emitting medium surrounding the black hole, the radiation is emitted by this accretion flow and detected in the plane of the distant observer. However, it is more convenient to calculate the photon trajectories backward in time, from the point of detection to the region around the black hole.

\begin{figure*}
\begin{center}
\includegraphics[type=pdf,ext=.pdf,read=.pdf,width=8.0cm]{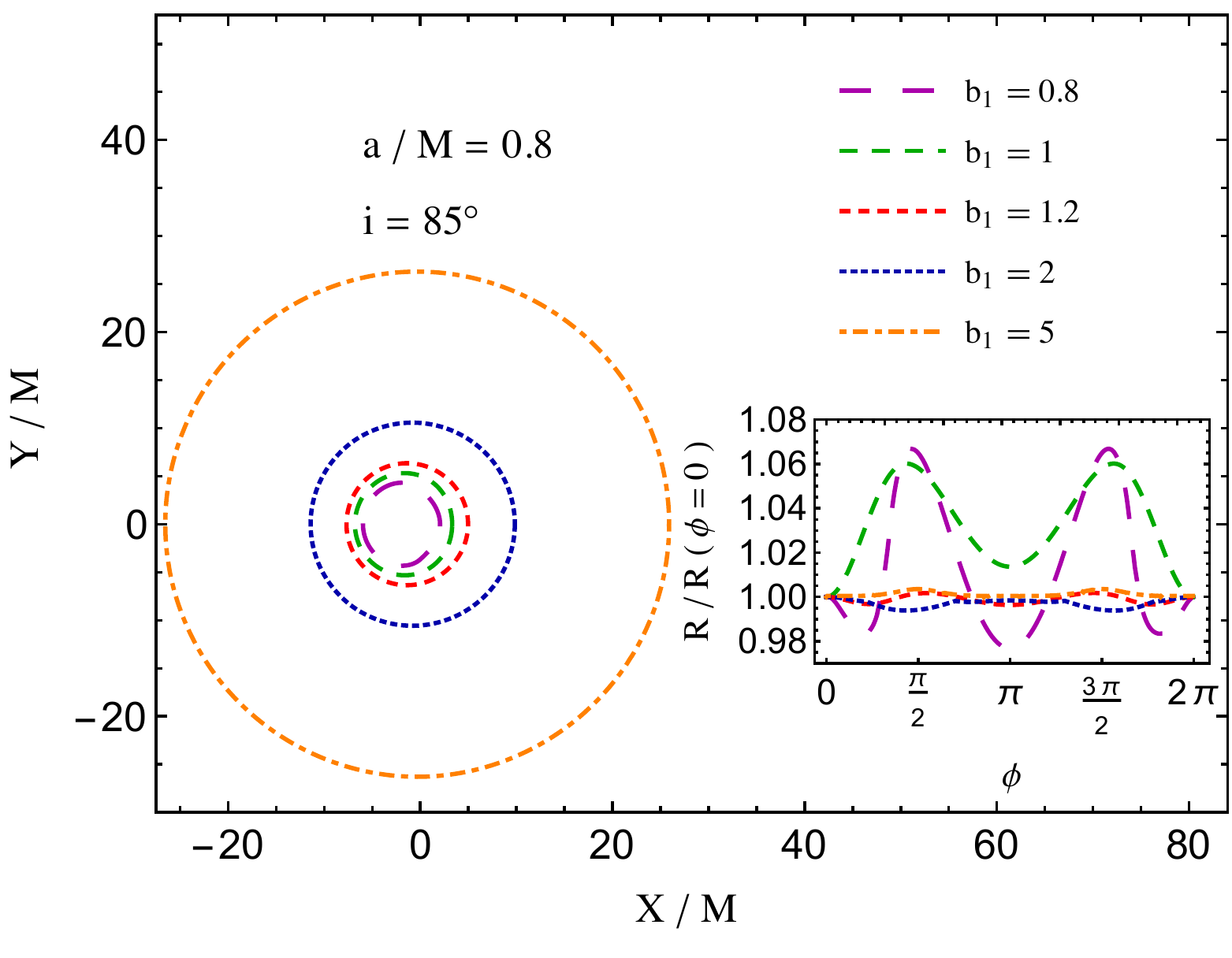}
\hspace{0.8cm}
\includegraphics[type=pdf,ext=.pdf,read=.pdf,width=8.0cm]{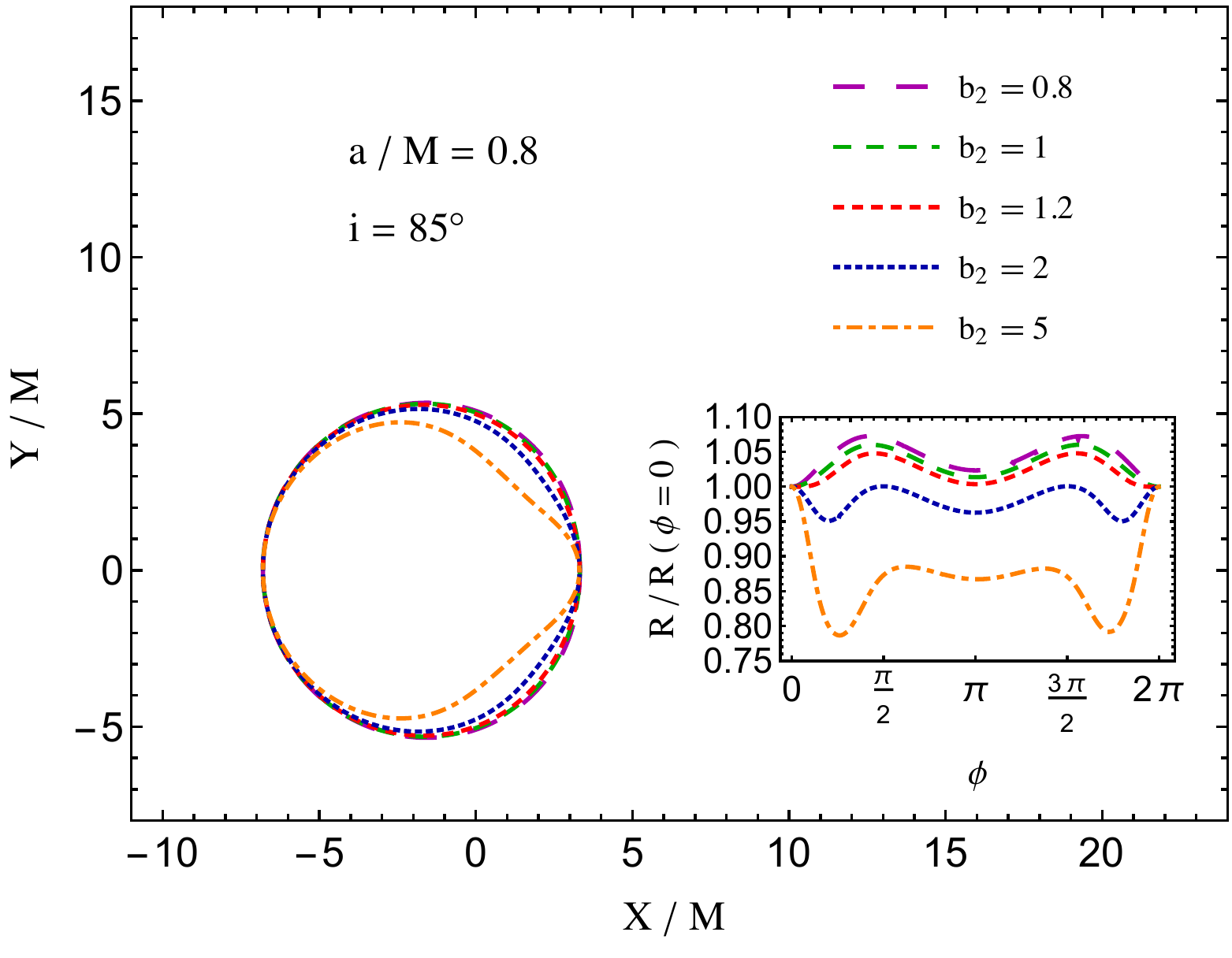}\\
\vspace{0.8cm}
\includegraphics[type=pdf,ext=.pdf,read=.pdf,width=8.0cm]{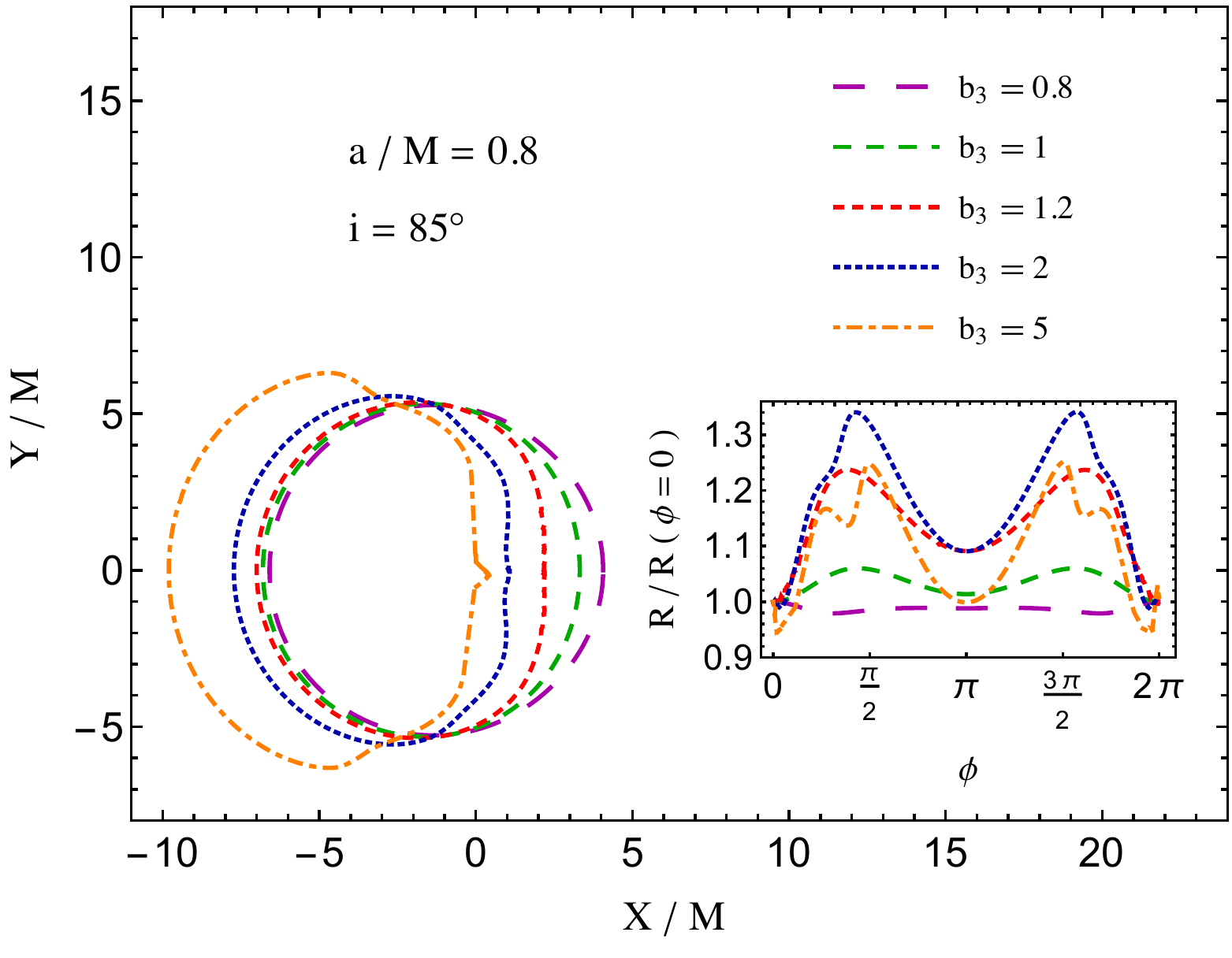}
\hspace{0.8cm}
\includegraphics[type=pdf,ext=.pdf,read=.pdf,width=8.0cm]{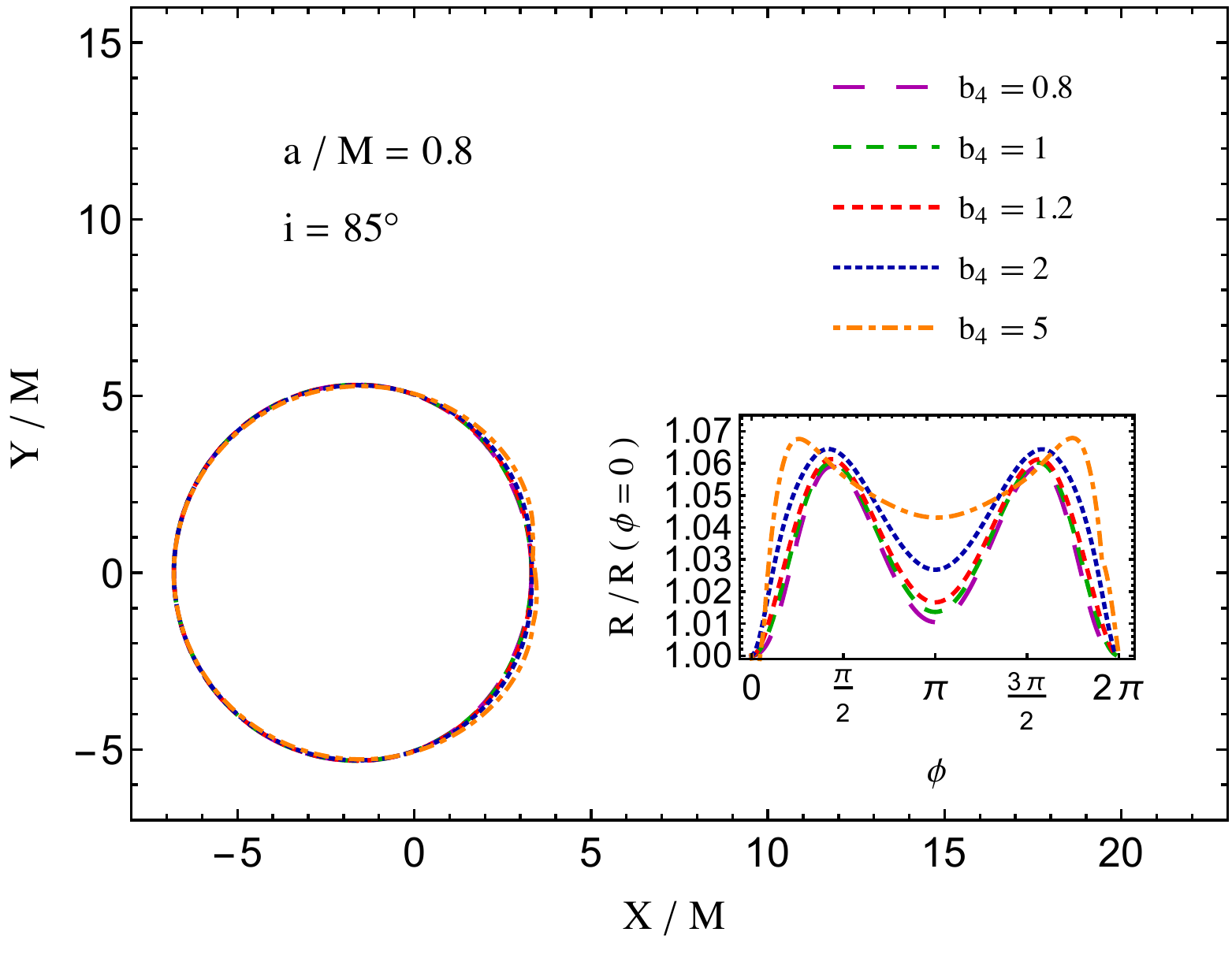}\\
\vspace{0.8cm}
\includegraphics[type=pdf,ext=.pdf,read=.pdf,width=8.0cm]{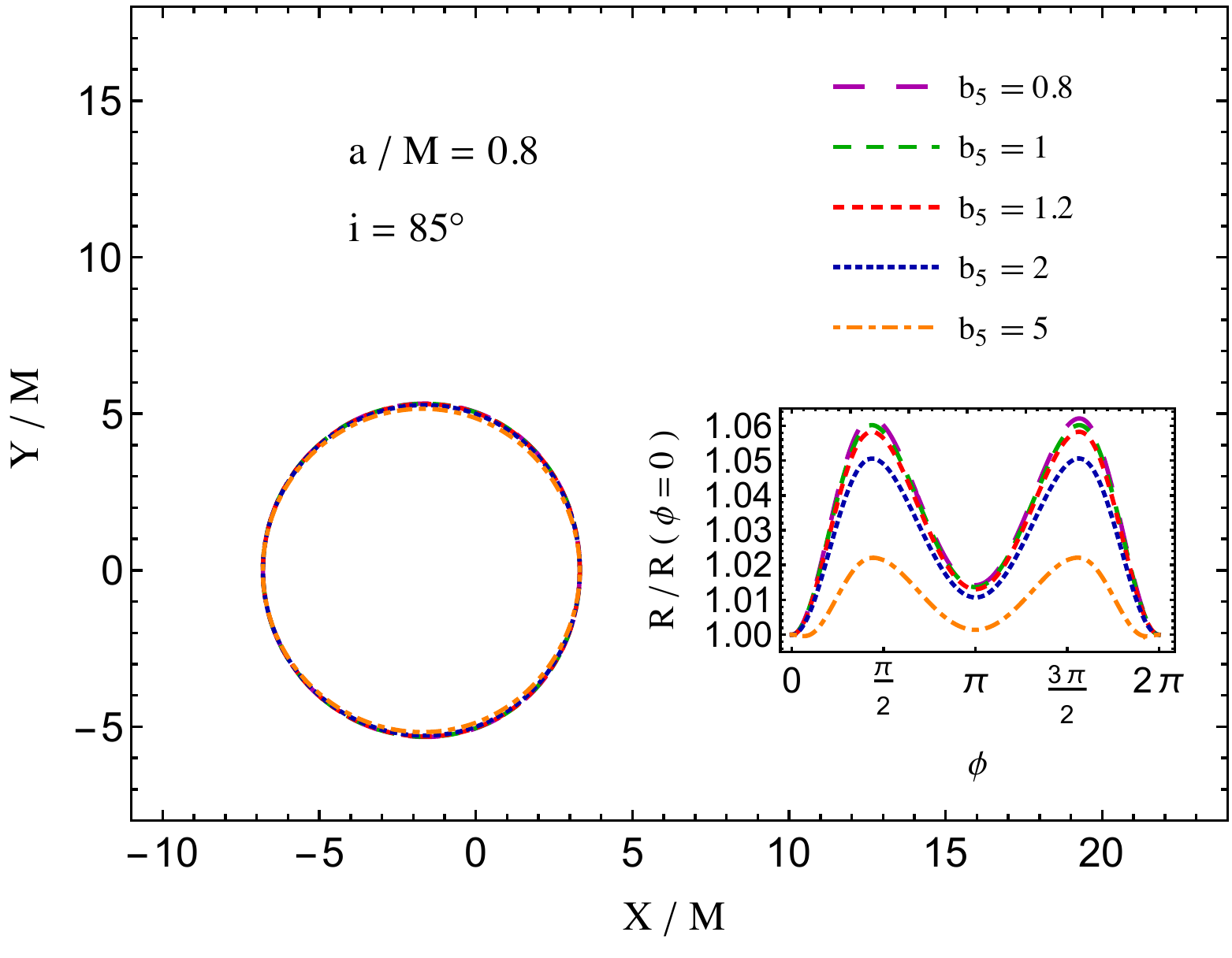}
\hspace{0.8cm}
\includegraphics[type=pdf,ext=.pdf,read=.pdf,width=8.0cm]{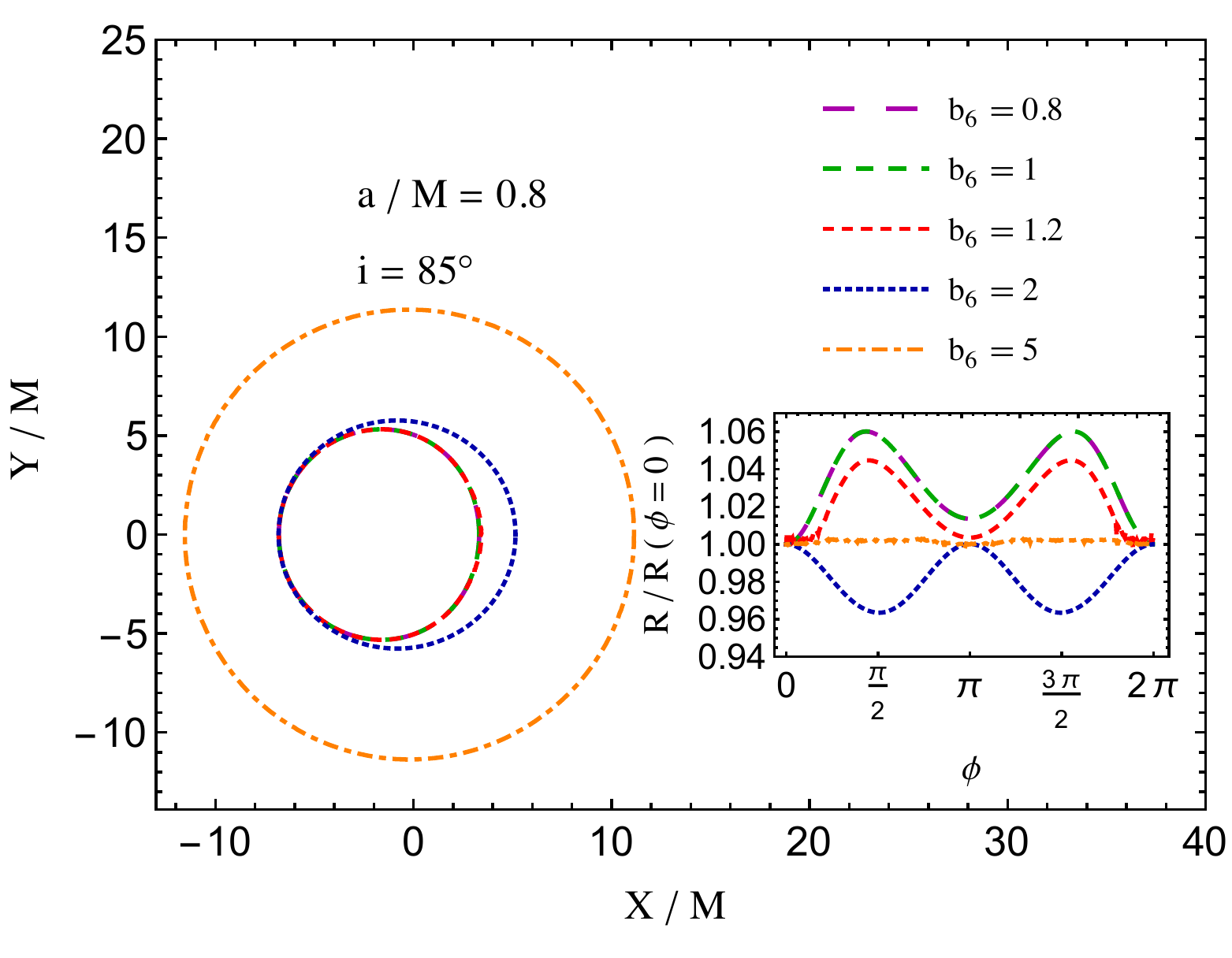} 
\end{center}
\caption{Impact of the parameters $b_1$ (top left panel), $b_2$ (top right panel), $b_3$ (central left panel), $b_4$ (central right panel), $b_5$ (bottom left panel), and $b_6$ (bottom right panel) on the shape of the shadow of a black hole. See the text for more details.}
\label{fig1}
\end{figure*}

\begin{figure*}
\begin{center}
\includegraphics[type=pdf,ext=.pdf,read=.pdf,width=8.0cm]{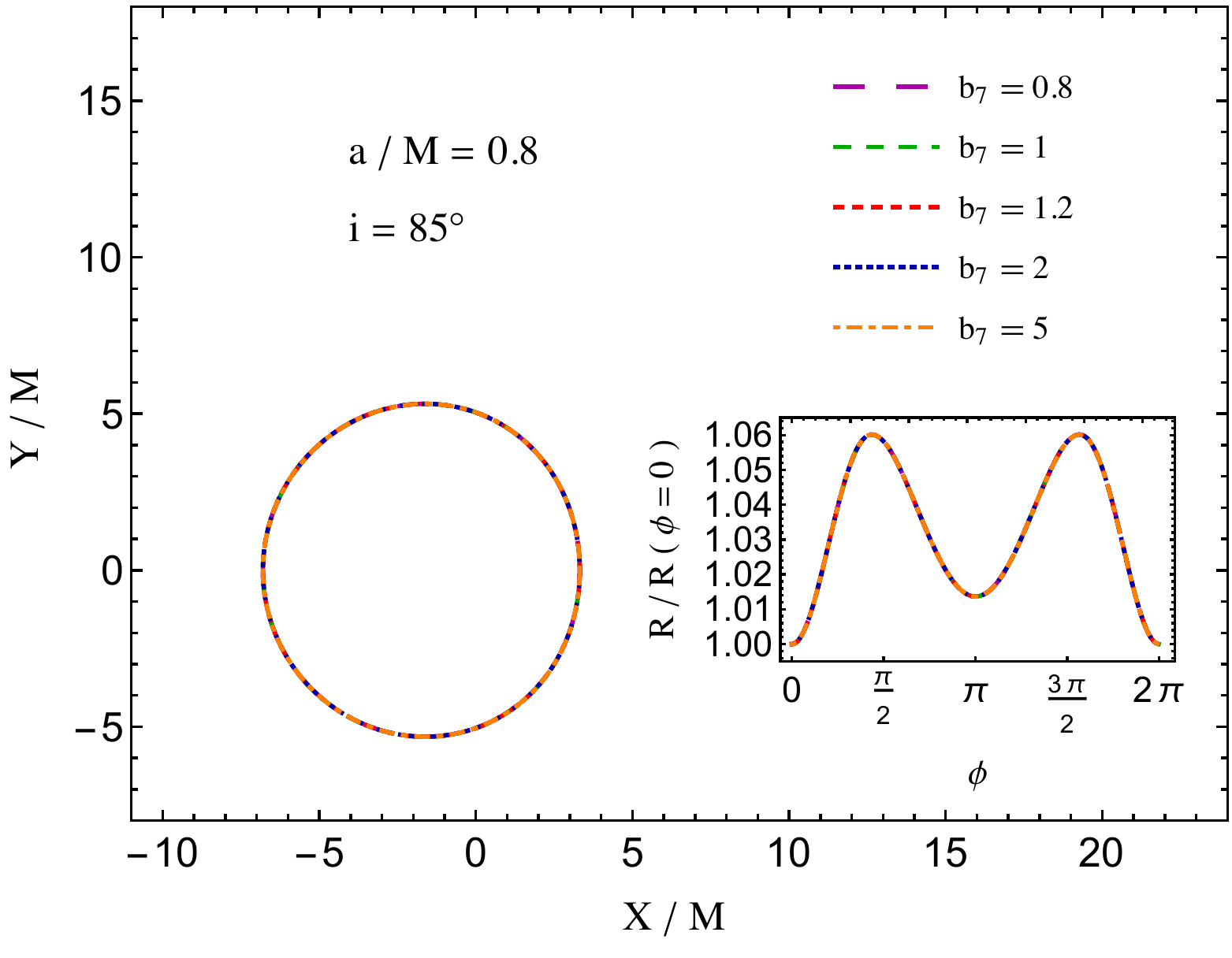}
\hspace{0.8cm}
\includegraphics[type=pdf,ext=.pdf,read=.pdf,width=8.0cm]{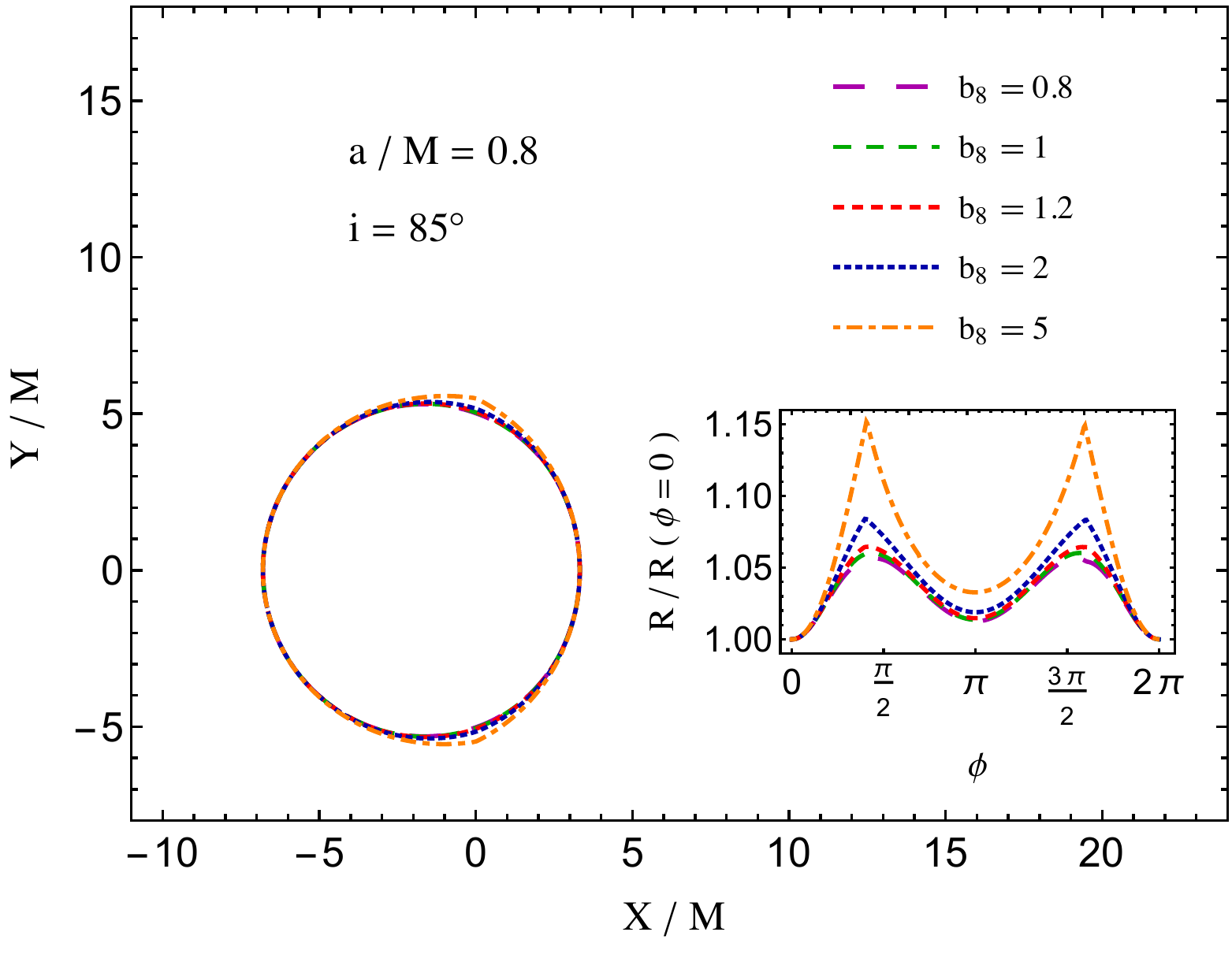}\\
\vspace{0.8cm}
\includegraphics[type=pdf,ext=.pdf,read=.pdf,width=8.0cm]{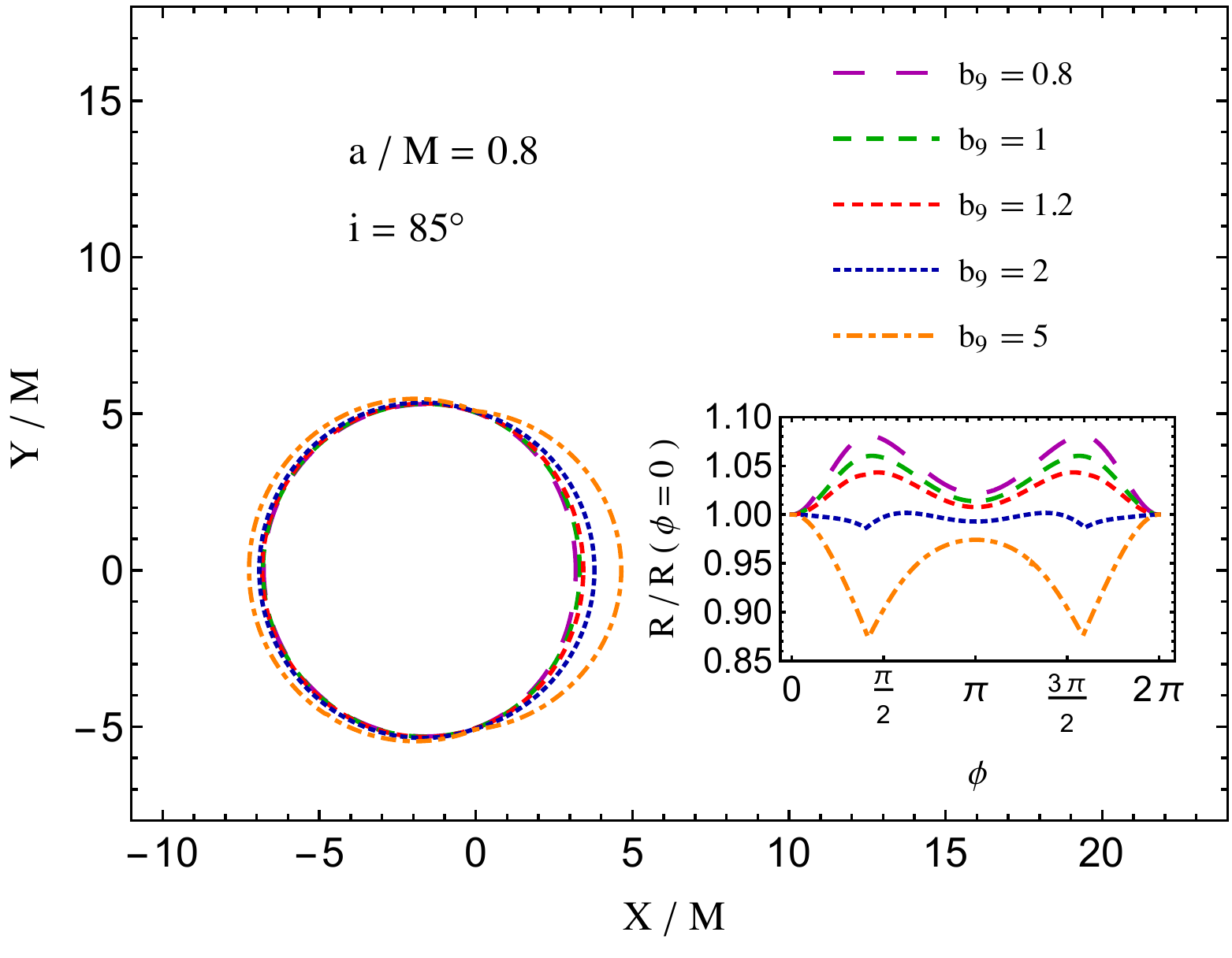}
\hspace{0.8cm}
\includegraphics[type=pdf,ext=.pdf,read=.pdf,width=8.0cm]{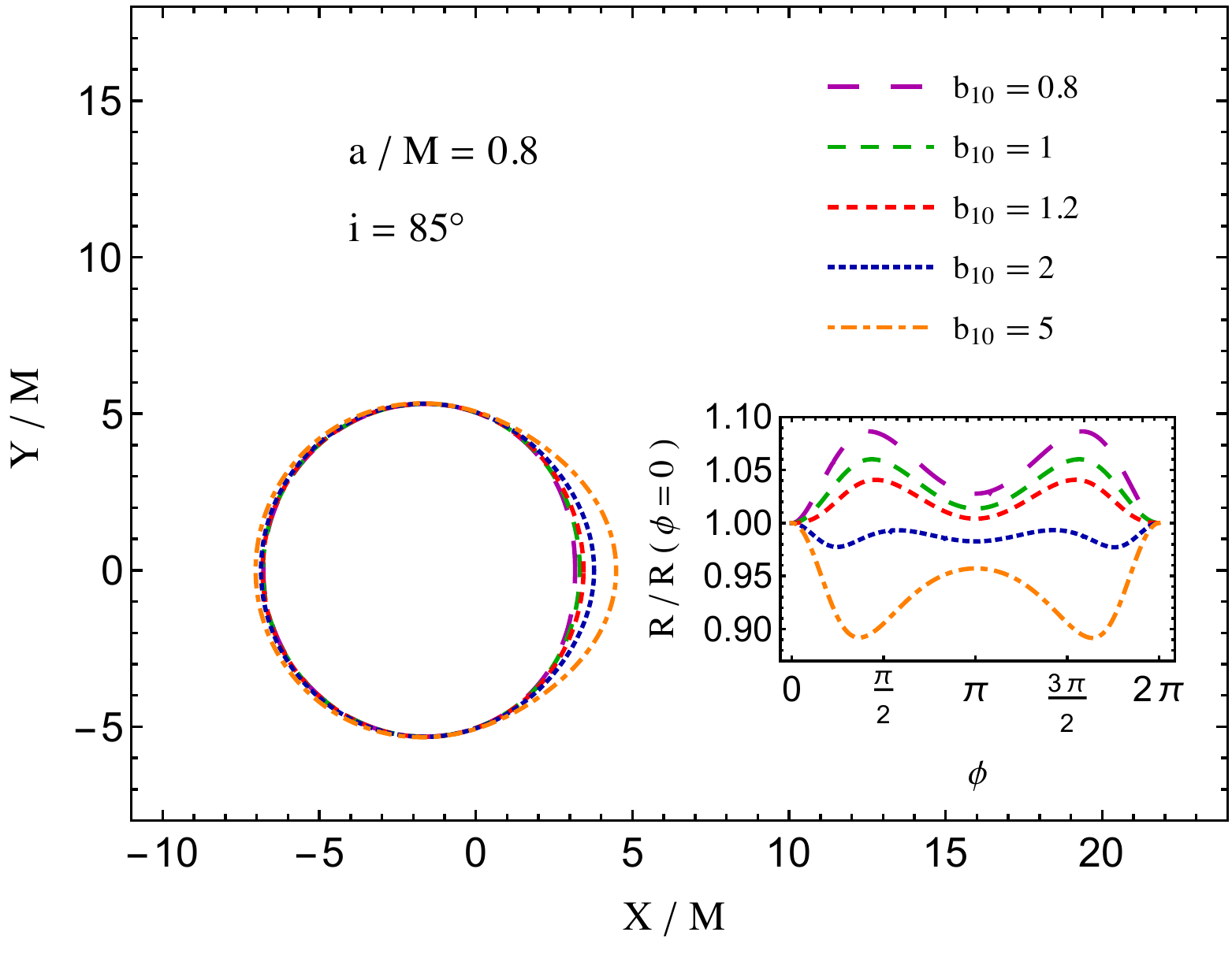}\\
\vspace{0.8cm}
\includegraphics[type=pdf,ext=.pdf,read=.pdf,width=8.0cm]{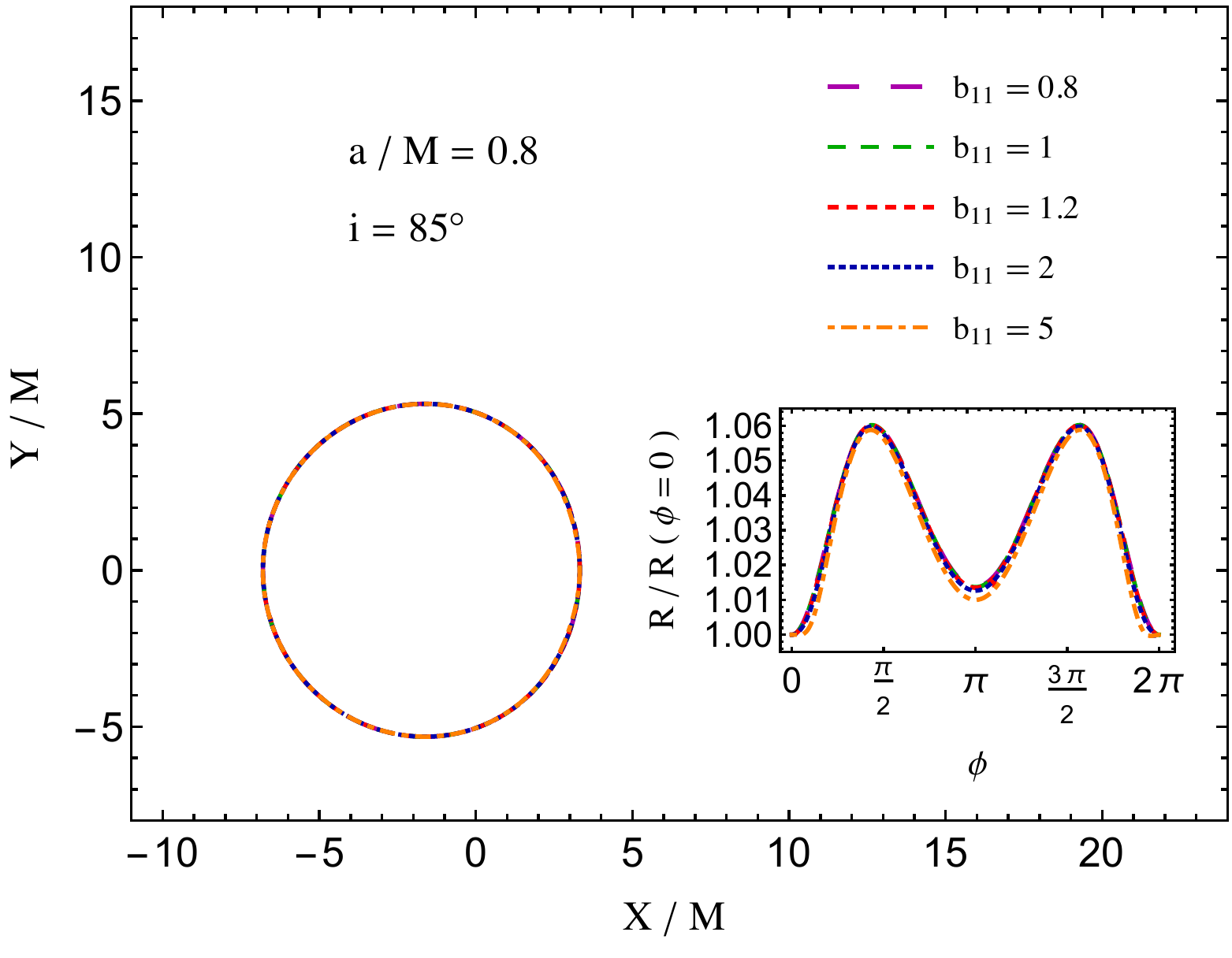} 
\end{center}
\caption{Impact of the parameters $b_7$ (top left panel), $b_8$ (top right panel), $b_9$ (central left panel), $b_{10}$ (central right panel), and $b_{11}$ (bottom panel) on the shape of the shadow of a black hole. See the text for more details.}
\label{fig2}
\end{figure*}

\begin{figure}
\begin{center}
\includegraphics[type=pdf,ext=.pdf,read=.pdf,width=7.0cm]{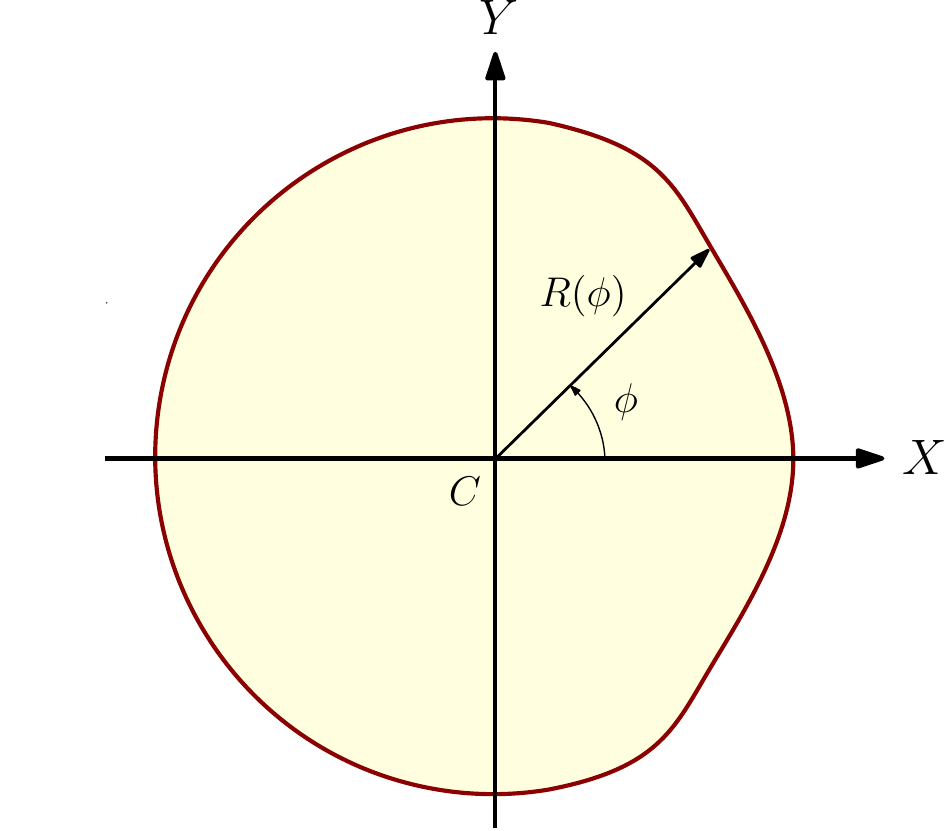}
\end{center}
\caption{The function $R(\phi)$ is defined as the distance between the center C and the boundary of the shadow at the angle $\phi$ as shown in this picture. See the text and Ref.~\cite{noi} for more details.}
\label{fig3}
\end{figure}

We have performed our calculations by changing the values of the coefficient $b_i$ and our results are shown in Figs.~\ref{fig1} and \ref{fig2}. For completeness, we have also studied the impact of $b_1$, $b_3$, and $b_6$. In all these plots, we show the case $a/M = 0.8$ and $i = 85^\circ$. The relatively high value of the spin parameter and the high inclination angle maximise the relativistic effects and therefore our choice represent a quite favourable case. These figures show both the shape of the black hole shadow (on the left in each panel) and the function $R(\phi)$ which was introduced in Ref.~\cite{noi} to characterise the shape of the shadow (see Ref.~\cite{rez} for a more sophisticated description). With reference to Fig.~\ref{fig3}, $R(\phi)$ is defined as the distance between the center of the shadow and the boundary of the shadow at the angle $\phi$. $R(0)$ is $R$ at $\phi = 0$ and the shape of the shadow is described by $R(\phi)/R(0)$.

\section{Discussion}

The coefficient $b_1$ (top left panel in Fig.~\ref{fig1}) affects the term responsible for the Newtonian limit. If we increases (decreases) the value of $b_1$, we are -- roughly specking -- considering a black hole with a larger (smaller) mass $M$. This is true for $b_1 > 1$. For very small values of $b_1$, the gravitational force is weak, but the black hole horizon is still in the same place because determined by the larger root of $g^{rr}$, namely by the equation
\be\label{eq-hor}
r^2 - 2 b_6 M r + b_7 a^2 = 0 \, .
\ee
If $b_1 < 1$, the usual photon capture radius determined by the the photon unstable orbit may be replaced by the event horizon.

The impact of $b_2$ on the black hole shadow is much smaller (top right panel in Fig.~\ref{fig1}). Only for large positive values it appreciably affects the boundary of the shadow. In other words, an observation may provide weak constraints on this coefficient and it is thus difficult to test the presence of the associated term in the Kerr metric.

The role of the coefficient $b_3$ is shown in the central left panel in Fig.~\ref{fig1}. Roughly speaking, as $b_1$ is the coefficient regulating the strength of the mass monopole moment, so $b_3$ control the strength of the leading term of the spin dipole moment. If $b_3 < 1$, the spin effects decrease and the shadow get more symmetric. For $b_3 > 1$, the difference between orbit with angular momentum parallel and antiparallel to the black hole spin increases. The force experienced by photons with angular momentum parallel to the black hole spin gets weaker and the photon capture radius on the right side of the shadow decreases. For very large $b_3$ ($b_3 = 5$ in Fig.~\ref{fig1}), the repulsive force of the spin is so strong that becomes dominant. The result is that there is very small or no capture sphere on the right side of the shadow: even if we fire a photon, close to the black hole the repulsive spin effect is stronger than the attractive mass term and the photon is scattered to infinity.

$b_4$ (central right panel in Fig.~\ref{fig1}) and $b_5$ (bottom left panel in Fig.~\ref{fig1}) have an extremely small impact on the shape of the shadow. Even when $b_4 = 5$ and $b_5 = 5$, deviations are very small. This means that the presence of such terms in the Kerr metric can unlikely be tested with the detection of the shadow of a black hole. It could be interesting to see if other approaches can do it, or otherwise $b_4$ and $b_5$ are always difficult to check.

$b_6$ (bottom right panel in Fig.~\ref{fig1}) is already constrained at the level of $10^{-5}$ by weak field tests in the Solar System and we cannot probably do better with tests in the strong gravity regime. $b_6$ determines also the position of the horizon via Eq.~(\ref{eq-hor}). If the value of $b_6$ is larger than 1, the radius of the event horizon may be larger than the standard photon capture sphere and the event horizon becomes the actual photon capture sphere, as shown in the bottom right panel in Fig.~\ref{fig1}. This is roughly equivalent to make $b_1$ sufficiently smaller than 1.

$b_7$ (top left panel in Fig.~\ref{fig2}) and $b_{11}$ (bottom panel in Fig.~\ref{fig2}) do not seem to produce any effect on the shape of the shadow. As in the case of $b_4$ and $b_5$, it would be interesting to see if other approaches can do this job or if these terms cannot be tested at all.

The impact of $b_8$ (top right panel in Fig.~\ref{fig2}), $b_9$ (central left panel in Fig.~\ref{fig2}), and $b_{10}$ (central right panel in Fig.~\ref{fig2}) on the boundary of the shadow is not large, but these coefficients produce characteristic features that are not present in the Kerr metric, for instance by changing the spin parameter or the viewing angle. In particular, $b_8$ and $b_9$ introduce two cusps in the function $R(\phi)$. $b_{10}$ makes the shadow oblate on the side of the photons with angular momentum parallel to the black hole spin.

Figs.~\ref{fig1} and \ref{fig2} can be better understood if we expand the line element in Eq.~(\ref{eq-proposal}) in $M/r$
\begin{widetext}
\be
ds^2 &=& - \left( 1 - \frac{2 b_1 M}{r}  + \frac{2 b_1 b_2 M a^2 \cos^2\theta}{r^3} + ... \right) dt^2 
- \left( \frac{4 b_3 M a \sin^2\theta}{r} 
- \frac{4 b_3 b_4 M a^3 \sin^2\theta \cos^2\theta}{r^3} + ... \right) dt d\phi \nonumber\\
&& + \left(1 + \frac{2 b_6 M}{r} 
+ \frac{ b_5 a^2 \cos^2\theta + 4 b_6^2 M^2 - b_7 a^2}{r^2} + ... \right)dr^2 
+ r^2 \left( 1 + \frac{b_8 a^2 \cos^2\theta}{r^2} + ... \right) d\theta^2 \nonumber\\ &&
+ \left( 1 + \frac{b_9 a^2}{r^2} + \frac{2 b_{10} M a^2 \sin^2\theta}{r^3}
- \frac{2 b_{10} b_{11} M a^4 \sin^2\theta \cos^2\theta}{r^5} + ... \right) r^2 \sin^2\theta d\phi^2 \, .
\ee
\end{widetext}
While the boundary of the shadow is determined by the photon orbits, where $M/r$ is not exactly a small parameter, this expansion provides a rough idea of the impact of each $b_i$. In particular, it makes evident that $b_4$ has a smaller impact than $b_3$, $b_5$ and $b_7$ have a smaller impact than $b_6$, and $b_{11}$ is more difficult to measure than $b_9$ and $b_{10}$.

\section{Summary and conclusions}

In this paper, we have proposed a new parametrization of the Kerr metric to test if the mass and the spin of a black hole deform the spacetime geometry as predicted by general relativity.

The common approach to perform tests of the Kerr metric is to consider a metric more general than the Kerr solution by introducing a number of extra terms. The latter can be turned on and off by their deformation parameters, which are adopted to quantify possible deviations from the Kerr geometry. The Kerr metric is recovered when all the deformation parameters vanish. Observations should measure the value of these deformation parameters to check whether they indeed vanish, as requested by the Kerr solution.

Here we have started from the Kerr metric in Boyer-Lindquist coordinates. We have introduced 11 ``Kerr parameters'' in front of any mass/spin term. These Kerr parameters should be 1 according to general relativity and may be different from 1 if the associated mass or spin term deforms the geometry in a different way. Some caution is in order here, because for some values of these parameters the spacetime can present pathological properties (naked singularities, closed time-like curves, etc.). In this way, we want to check whether each mass and spin term appearing in the Kerr metric make the spacetime geometry curved with the strength predicted by general relativity.

We have applied our proposal to the case of the shadow of a black hole. Our conclusions are as follows. The coefficients $b_1 = 1$ to recover the correct Newtonian limit. $b_3$, and $b_6$ are already constrained by tests in weak gravitational field and we can unlikely do better with tests in the strong gravity regime. $b_2$, $b_8$, $b_9$, and $b_{10}$ leave some small signatures in the boundary of the shadow. At least in principle, very accurate observations may test these coefficients. $b_4$, $b_5$, $b_7$, and $b_{11}$ do not seem to leave any specific signature. Their impact is small or absent, which means that even an extremely accurate measurement of the exact shape of the shadow cannot test the associated terms in the Kerr metric.


\begin{acknowledgments}
This work was supported by the NSFC grants No.~11305038 and No.~U1531117, the Shanghai Municipal Education Commission grant for Innovative Programs No.~14ZZ001, and the Thousand Young Talents Program. M.G.-N. acknowledges also support from China Scholarship Council (CSC), grant No.~2014GXZY08. C.B. acknowledges also support from the Alexander von Humboldt Foundation.
\end{acknowledgments}


\end{document}